\documentstyle[12pt]{article}
\begin{document}

\title{Lax representation for two--particle dynamics splitting on
  two tori} \author{ V Z Enolskii${}^{1)}$ and M Salerno ${}^{2)}$
\\ ${}^{1)}$Department of Theoretical Physics \\ Institute of
Magnetism \\of National Academy of Sciences of Ukraine,\\ Vernadsky
str.  36, Kiev-680, 252142, Ukraine\\
${}^{2)}$Istituto Nazionale di Fisica della Materia and\\
Department of Theoretical Physics Institute of
Physics,\\ University of Salerno, Baronissi
I-84100 , Italy}
\date{\empty}

\maketitle

\abstract{
\noindent
Lax representation in terms of $2\times 2$ matrices is constructed
for separable multiply--periodic systems splitting on two tori.
Hyperelliptic Kleinian functions and their reduction to elliptic
functions are used. }

\vskip 2cm
{\bf Introduction.} Completely integrable systems with two degree
of freedoms and with dynamics splitting on two tori have been
largely investigated during the past years as  examples of
separable multiply-periodic systems. The list of such  systems
includes the well known integrable cases of the H\'enon-Heiles
system \cite{rgg93}, several integrable cases of quartic
potentials \cite{rrg94},  the motion of a particle in the Coulomb
potential and in external uniform field, the Chaplygin top
\cite{cha54}, etc. A Lax representation for these systems can be
readily constructed in terms of a direct product of Lax operators
\cite{rgg93}, one for each splitting tori, as first proposed in
ref.\cite{vanh92}. This approach leads for a system of two
particles to $4\times 4$ Lax representations (see e.g.
\cite{rgg93,rrg94}) this making the quantization of the
above systems much more difficult to perform.  In order to
simplify the quantum problem it would be more convenient to use
Lax representations in terms of $2 \times 2$ matrices. The problem
of the existence of such representations for the above systems is
still open.

The aim of the present paper is to show how
to construct for dynamics splitting on two tori, Lax
representations in terms of $2\times2$ matrices.  The main idea is
to use hyperelliptic curve of genus two, which is a $N$--sheeted
cover of two given elliptic curves.  Such covers are known to
exist for any $N>1$ and for arbitrary tori (see
e.g.\cite{ba97,kr03,bbeim94}).  It is
clear that if the hyperelliptic curve is associated with an
hamiltonian system for which a $2\times2$ Lax representation is
known, one can readily construct a similar representation for the
two tori dynamics simply by using the transformation  induced by
the covering.  To illustrate this approach we take as working
example the integrable cases of the
H\'enon--Heiles system  \cite{fo91}.  The possibilities of
generalization of this approach to system with more then two
degrees of freedom is briefly discussed at the end of the
paper.

{\bf Reduction.} Consider the hyperelliptic curve $V=(y,z)$ of
genus two, \begin{equation} y^2=4z^5+\lambda_4x^4
+\lambda_3x^3+\lambda_2x^2+\lambda_1x+\lambda_0 \label{curveg}
\end{equation}
with $\lambda_i\in {\bf C}$ chosen in such a way that
(\ref{curveg}) takes the form
\begin{equation} w^2=z(z-1)(z-\alpha)(z-\beta)(z-\alpha\beta)\;.
\label{curver}\end{equation}
The curve (\ref{curver}) gives a two-sheeted covering of two tori
$\pi_{\pm}: V = (w,z) \rightarrow E_{\pm} = (\eta_{\pm}, \xi_{\pm})$,
\begin{equation}
\eta_{\pm}^2=\xi_{\pm}(1-\xi_{\pm})(1-k_{\pm}^2\xi_{\pm})\;,
\end{equation} with Jacobi moduli \begin{equation}
k_{\pm}^2=-\frac{(\sqrt{\alpha}\mp\sqrt{\beta})^2}{(1-\alpha)(1-\beta)}\;.
\label{jacmod} \end{equation}
Equation (\ref{jacmod}) can be inverted as
\begin{equation}
\alpha+\beta=2\frac{k_+^2+k_-^2}{({k_+^{\prime}}-{k_+^{\prime}})^2}\;,\quad
\alpha\beta=\left(\frac{k_+^{\prime}+k_-^{\prime}}{k_+^{\prime}
-k_+^{\prime}}\right)^2\;,
\label{jacmodinv}\end{equation}
where $k_{\pm}^{\prime}$ are additional Jacobian moduli,
$k_{\pm}^2+{k_{\pm}^{\prime}}^2=1$.
Explicitly, the covers $\pi_{\pm}$ are given by
\begin{eqnarray}
\eta_{\pm}&=&-\sqrt{(1-\alpha)(1-\beta)}\quad \frac{z\mp
\sqrt{\alpha\beta}} {(z-\alpha)^2(z-\beta)^2}\; w\;,\label{r1}\\
\xi_+&=&\xi_-=\frac{(1-\alpha)(1-\beta)z}{(z-\alpha)(z-\beta)}\;.\label{r2}
\end{eqnarray}
\noindent
Let $(y_1,x_1), (y_2,x_2)$ be arbitrary points on a symmetric
degree $V\times V$. The Jacobi inversion problem is the problem of
finding this point as a function ${\bf u}=(u_1,u_2)$ from the
equations
 \begin{eqnarray} \int_{x_0}^{x_1} \frac{dz}{w}+
\int_{x_0}^{x_2} \frac{dz}{w}=u_1\;, \label{j1}\\
\int_{x_0}^{x_1}\frac{zdz}{w} +
\int_{x_0}^{x_2} \frac{zdz}{w}=u_2\;.\label{j2}
\end{eqnarray}   We write \begin{eqnarray} \int_{x_0}^{x_1}
\frac{z-\sqrt{\alpha\beta}}{w}dz +\int_{x_0}^{x_2}
\frac{z-\sqrt{\alpha\beta}}{w}dz=u_+\;,\label{j11}\\
\int_{x_0}^{x_1} \frac{z+\sqrt{\alpha\beta}}{w}dz
+\int_{x_0}^{x_2}
\frac{z+\sqrt{\alpha\beta}}{w}dz=u_-\;.\label{j22} \end{eqnarray}
with
\begin{equation}
u_{\pm}=-\sqrt{(1-\alpha)(1-\beta)}\quad
(u_2\mp\sqrt{\alpha\beta}u_1)\;.  \end{equation} We can reduce the
hy\-per\-el\-lip\-tic in\-teg\-rals in  (\ref{j11},\ref{j22}) to
el\-lip\-tic ones by using the formula \begin{equation}
\frac{d\xi_{\pm}}{\eta_{\pm}}=-\sqrt{(1-\alpha)(1-\beta)}\quad
(z\mp\sqrt{\alpha\beta})\frac{dz}{w}\;.\label{r3}
\end{equation}
Let us introduce the coordinates  (see \cite{hu94})
\begin{eqnarray}
X_1&=& {\rm sn}(u_{+},k_{+}){\rm sn}(u_{-},k_{-})\;,\cr
X_2&=& {\rm cn}(u_{+},k_{+}){\rm cn}(u_{-},k_{-})\;,\label{darb}\\
X_3&=& {\rm dn}(u_{+},k_{+}){\rm dn}(u_{-},k_{-})\;,\nonumber
\end{eqnarray}
where ${\rm sn}(u_{\pm},k_{\pm})$,  ${\rm cn}(u_{\pm},k_{\pm})$,
and ${\rm dn}(u_{\pm},k_{\pm})$ denote usual Jacobi elliptic
functions\cite{ba55}.
Applying the addition theorem for Jacobi elliptic functions,
\begin{eqnarray} {\rm
sn}(u_1+u_2,k)&=&\frac{s_1^2-s_2^2}{s_1c_2d_2-s_2c_1d_1}\;, \cr
{\rm
cn}(u_1+u_2,k)&=&\frac{s_1c_1d_2-s_2c_2d_1}{s_1c_2d_2-s_2c_1d_1}\;,\cr
{\rm
dn}(u_1+u_2,k)&=&\frac{s_1d_1c_2-s_2d_2s_1}{s_1c_2d_2-s_2c_1d_1}\;, \nonumber \end{eqnarray} where
$s_i={\rm sn}(u_i,k),\;
c_i={\rm cn}(u_i,k), \;d_i={\rm dn}(u_i,k) $, $i=1,2$,  we can
write Eq.s (\ref{darb}) in the form \begin{eqnarray}
X_1&=&-\frac{(1-\alpha)(1-\beta)(\alpha\beta+\wp_{12})}
{(\alpha+\beta)(\wp_{12}-\alpha\beta)+\alpha\beta\wp_{22}+\wp_{11}}\; ,
\label{x1}\\
X_2&=&-\frac{(1+\alpha\beta)(\alpha\beta-\wp_{12})-\alpha\beta\wp_{22}
-\wp_{11}}
{(\alpha+\beta)(\wp_{12}-\alpha\beta)+\alpha\beta\wp_{22}+\wp_{11}}\quad,
\label{x2}\\ X_3&=&\frac{\alpha\beta\wp_{22}-\wp_{11}}
{(\alpha+\beta)(\wp_{12}-\alpha\beta)+\alpha\beta\wp_{22}+\wp_{11}}\;.
\label{x3}\end{eqnarray}
Here $\wp_{ij}$ are Kleinian $\wp$--functions
which solve the Jacobi inversion problem and are
expressed in terms of $(y_1,x_1),(y_2,x_2)$ as follows
\begin{eqnarray*}
\wp_{22}=x_1+x_2,\quad\wp_{12}=-x_1x_2\;,\quad
\wp_{11}=\frac{F(x_1,x_2)-2y_1y_2}{4(x_1-x_2)^2}\cr
\end{eqnarray*}
and
\begin{equation}
F(x_1,x_2)=\sum_{k=0}^{k=2}x_1^kx_2^k(2\lambda_{2k}
+\lambda_{2k+1}(x_1+x_2))
\end{equation}
with $\lambda$'s calculated from (\ref{curver}).
The Kleinian $\wp$--functions are known to be a natural
generalization of the Weierstrass elliptic functions and can
then be expressed through second logarithmic derivative of the Kleinian
$\sigma$--function, $$\wp_{ij}({\bf u})
=-\frac{\partial^2\;{\rm ln}\sigma({\bf u})}{\partial u_i\partial
u_j}\;,\quad i,j=1,2, $$
(for details see \cite{ba97,bel96a}). The three
functions $\wp_{22},\wp_{12},\wp_{11}$ are algebraically dependent
and are coordinates for the so called Kummer surface which
is a quartic surface in ${\bf C}^3$.
For later convenience we remark that the
formulae (\ref{x1}-\ref{x3}) can be inverted as
\begin{eqnarray}
\wp_{11}&=&(B-1)\frac{A(X_2+X_3)-B(X_3+1)}
{X_1+X_2-1}\;,\label{wp11}\\
\wp_{12}&=&(B-1)\frac{1+X_1-X_2}{X_1+X_2-1}\;,\label{wp12}\\
\wp_{22}&=&\frac{A(X_2-X_3)+B(X_3-1)}
{X_1+X_2-1}\;,\label{wp22}\end{eqnarray}
where $A=\alpha+\beta$, $B=1+\alpha\beta$.

{\bf Lax representation.} Let us consider the following equations
for the four--indexed functions $\wp$:  \begin{eqnarray}
\wp_{2222}&=&6\wp_{22}^2+4\wp_{12}+\lambda_4\wp_{22}+\frac{1}{2}
\lambda_3\;,\label{kdv1}\\
\wp_{1222}&=&6\wp_{22}\wp_{12}-2\wp_{11}+\lambda_4\wp_{12}\;,\label{kdv2}
\end{eqnarray}
with $\lambda_3$ and $\lambda_4$ arbitrary . The first equation,
after $u_2$ differentiation, is the standard
KdV equation written with respect to the
function $\wp_{22}$ while the second equation represents the stationary
flow for the two gap KdV--solution ($\wp_{22}$) of the
 third vector field  of the KdV hierarchy.
As well known equations (\ref{kdv1},\ref{kdv2}) can be
written in the Lax form \cite{zmnp80},
\begin{equation}
\frac{\partial L}{\partial t}=[M,L],\quad
L=\left(\begin{array}{cc}V&U\\W&-V\end{array}\right),\quad
M=\left(\begin{array}{cc}0&1\\Q&0\end{array}\right)\;.  \label{lax}
\end{equation}
Here we take  the elements of the matrices $L$ and $M$ to
be polynomials in $x$ of the form
\begin{eqnarray}
U&=&x^2-\wp_{22}x-\wp_{12},\label{u}\\
V&=&-\frac12\frac{\partial U}{\partial u_2}\;,\label{v}\\
W&=&-\frac12\frac{\partial^2 U}{\partial u_2^2}+UQ\;,\label{w}\\
Q&=&x+2\wp_{22}+\frac14\lambda_4\;.\label{q}
\end{eqnarray}
The discriminant curve ${\rm det}\; (L -yE)=0,$ ($E$ is the
$2\times2$ unit matrix) has then the form of Eq. (\ref{curveg}) with
$\lambda_4,\lambda_3, \lambda_0$ arbitrary and $\lambda_2$,
$\lambda_1$ chosen as the level set of the integrals of motion:
\begin{eqnarray}
-\lambda_2&=&-\wp_{222}^2+4\wp_{11}+\lambda_3\wp_{22}+4\wp_{22}^3
\cr&+&4\wp_{12}\wp_{22}
+\lambda_4\wp_{22}^2\;,\label{int1}\\
-\frac12\lambda_1&=&-\wp_{222}\wp_{221}+2\wp_{12}^2-2\wp_{11}\wp_{22}
+{1\over2}\lambda_3\wp_{12}\label{int2}\\
&+&4\wp_{12}\wp_{22}^2+\lambda_4\wp_{12}\wp_{22}\;.
\nonumber\end{eqnarray}
The following proposition represents the main result of the paper.

{\bf Proposition}
{\it Let
\begin{eqnarray}
U&=&x^2-\frac{A(X_2-X_3)+B(X_3-1)}{X_1+X_2-1}\;x\cr
&+&(B-1)\frac{X_1-X_2+1}{X_1+X_2-1}\;,\label{uu}\\
Q&=&x+2\frac{A(X_2-X_3)+B(X_3-1)}{X_1+X_2-1}+A+B\;,\label{qq}
\end{eqnarray}
where $X_i$ are the coordinates given in (\ref{darb}) and
$A=\alpha+\beta,B=\alpha\beta+1$ are expressed in terms of Jacobian
moduli $k_{\pm}$ according to (\ref{jacmodinv}). Then the Lax equation
(\ref{lax}) is equivalent to the equations for Jacobi elliptic
functions,
\begin{equation}
\frac{d}{d u_{\pm}}{\rm sn}(u_{\pm};k_{\pm})=
\sqrt{(1-{\rm sn}^2(u_{\pm};k_{\pm})(1-k^2_{\pm}{\rm
sn}^2(u_{\pm};k_{\pm}))}\;. \label{ja} \end{equation} }
To prove this statement one can expand
${\rm sn}(u_{\pm};k_{\pm})$
around $u_{\pm}=0$ to obtain from (\ref{lax}) the equation
(\ref{ja}) with the superscripts '$\pm$'. We remark that a direct
substitution of (\ref{wp11},\ref{wp12},\ref{wp22}) into the equations of
motion (\ref{kdv1},\ref{kdv2}) would be quite
involved even for symbolic calculations on a computer.

{\bf An example: Lax representation for the integrable cases of
the H\'enon--Heiles system.} Let us apply the above result to the
integrable cases of the H\'enon--Heiles system (see
e.g.\cite{fo91}).  One of them (the case ({\bf ii}) in the
terminology of \cite{fo91}) is isomorphic to the fifth--order
stationary KdV flow, this giving a
Lax representation in terms of $2\times2$ matrices. The other
two cases-- the
cases ({\bf i}) and
({\bf iii})-- are isomorphic to fifth stationary flows of
respectively Sawada--Kotera and Kaup-Kupershmidt equations. They
both lead to a Lax representations in terms of $3\times3$
matrices\cite{fo91}. The $4\times4$ Lax representation is
derived in \cite{rgg93}. Let us show how to construct $2\times 2$
Lax representation for the cases ({\bf i}),({\bf iii}).

Consider first the integrable case ({\bf i}).  The Hamiltonian $H$
and second integral of motion $K$ have the form \begin{eqnarray}
H&=&\frac12p_1^2+\frac12p_2^2+q_1q_2^2+\frac13q_1^3+a(q_1^2+q_2^2)\;,
\label{i1}\\
K&=&p_1p_2+\frac13q_2^3+q_2q_1^2+2aq_1q_2\;.\label{i2}
\end{eqnarray}
The hamiltonian system is separated in Cartesian coordinates,
$q_{1,2}=\tilde Q_1\pm \tilde Q_2$ , $p_{1,2}=\tilde P_1\pm \tilde
P_2$ and the dynamics is splitting to two tori
\begin{eqnarray}
\tilde P_1^2&=&-\frac43 \tilde Q_1^3-2a\tilde Q_1^2
+\frac12(\tilde H+\tilde K)\;,\nonumber\\
\tilde P_2^2&=&-\frac43 \tilde Q_2^3-2a\tilde Q_2^2
+\frac12(\tilde H-\tilde K)\;, \label{anharm}
\end{eqnarray}
where
$\tilde H= \tilde P_1^2 +\tilde P_2^2 +\frac43\tilde
Q_1^3 +\frac43\tilde Q_3^3 +2a(\tilde Q_1^2+\tilde Q_2^2),
\tilde K =\tilde P_1^2 -\tilde P_2^2 +\frac43\tilde
Q_1^3 -\frac43\tilde Q_3^3+ 2a(\tilde Q_1^2 -\tilde
Q_2^2)$. By passing from (\ref{anharm}) to
the standard form of the elliptic curve (\ref{ja}) we find
\begin{equation}
\wp^{\pm}\left(\frac{it}{\sqrt{3}}\right)=\frac12(q_1(t)\pm q_2(t)+a)
\end{equation}
with $\wp^{\pm}$ standard Weierstrass elliptic functions with moduli
$e_i^{\pm}$, $i=1,2,3$ satisfying the equations
\begin{equation}
4e_1^{\pm}e_2^{\pm}e_3^{\pm}=a^3-\frac{3}{2}(\tilde H\pm\tilde
K)\;,\quad
8(e_1^{\pm}e_2^{\pm}+e_1^{\pm}e_3^{\pm}+e_2^{\pm}e_3^{\pm})+3\frac{a^2}{2}=0\;.
\end{equation}
The Lax representation (\ref{lax}) is then valid for the system with
\begin{eqnarray}
X_1&=&\sqrt{\frac{2e_1^+-2e_3^+}{q_1+q_2+a-2e_3^+}}
\sqrt{\frac{2e_1^--2e_3^-}{q_1-q_2+a-2e_3^-}}\;. \cr
X_2&=&\sqrt{\frac{q_1+q_2+a-2e_1^+}{q_1+q_2+a-2e_3^+}}
\sqrt{\frac{q_1-q_2+a-2e_1^-}{q_1-q_2-2e_3^-}}\;. \cr
X_3&=&\sqrt{\frac{q_1+q_2+a-2e_2^+}{q_1+q_2+a-2e_3^+}}
\sqrt{\frac{q_1-q_2+a-2e_2^-}{q_1-q_2+a-2e_3^-}}\;.\nonumber
\end{eqnarray}
As shown in ref.\cite{sel94}, the integrable case {\bf (iii)} is
linked to case {\bf (i)} by means of a
canonical transformation. The corresponding $2\times2$
Lax representation can be then derived from the one of case {\bf
(i)} by means of this transformation.

{\bf Concluding remarks.} In closing this paper we make the
following remark.  Equip the curve by the canonical basis of
cycles $A_1,A_2,B_1,B_2$ and normalize the holomorphic
differentials $dv_i=(c_{i1}+zc_{i2})dz/w(z)$, $i=1,2$ in such a
way that the Riemann matrix $\Omega$ has the following form
\begin{equation}
\Omega
=\left(\begin{array}{cccc}
\oint_{A_1}dv_1&\oint_{A_2}dv_1&\oint_{B_1}dv_1&\oint_{B_2}dv_1\\
\oint_{A_1}dv_2&\oint_{A_2}dv_2&\oint_{B_1}dv_2&\oint_{B_2}dv_2
\end{array}\right)
=\left(\begin{array}{cccc}
1&0&\tau_{11}&\tau_{12}\\
0&1&\tau_{12}&\tau_{22}
\end{array}\right)   \end{equation}

It is known (see e.g. \cite{kr03,hu94,bbeim94}),
that the curve (\ref{curveg}) cover $N$--sheetedly two
tori if and only if the Riemann matrix $\Omega$ can be transformed
by some linear transformation of the basis cycles to the form
\[ \tau=\left(\begin{array}{cc}
\tau_{11}&{1\over N}\\{1\over N}&\tau_{22}\end{array}\right),\]
where the positive integer $N$ is also called a Picard number. The
condition for the matrix $\tau$ to be transformed to the form given
above is that $\tau$ belongs to the Humbert surface $H_{N}$
\begin{eqnarray}
 H_{N} &=&\left\{\alpha \tau_{11} + \beta \tau_{12} + \gamma
\tau_{22} + \delta (\tau^{2}_{12} - \tau_{11}\tau_{22}) +\varepsilon
=0,\right.\nonumber\\&& \left.  \alpha,\beta , \gamma , \delta ,
\varepsilon  \in  {{\bf   Z}}, \quad \beta^{2} - 4(\alpha \gamma  +
\varepsilon \delta)=N^2\right\}.\nonumber \end{eqnarray}

The case considered in this paper corresponds,
among the infinite
transformations of $N$--th order which permit to reduce a dynamics
of two particle system associated with $N$--sheeted covering of tori,
just to the case $N=2$.
It is clear, however, that the above  analysis
can be extended to curves of high genus.

These arguments were used in \cite{bbeim94} to describe elliptic
potentials of the Schr\"odinger equation, which also studied in
the frameworks of spectral theory
\cite{gw95,gw95a,gw95b}.

{\bf Acknowledgement}. The authors are grateful to V.Kuznetsov
who attracted their attention to the problem discussed.  The
researches described in this publication were supported in part by
grant no.  U44000 (VZE) from the ISF and INTAS grant no.
93--1324~(VZE and MS).

\end{document}